\documentclass[conference,letterpaper]{IEEEtran}

\usepackage[utf8]{inputenc} 
\usepackage[T1]{fontenc}
\usepackage{url}
\usepackage{ifthen}
\usepackage{cite}

\usepackage{graphicx}
\usepackage{hyperref}
\usepackage{amsmath}
\usepackage{amssymb}
\usepackage{bm}
\usepackage{bbm}
\usepackage{algorithmicx}
\usepackage{algorithm}
\usepackage{algpseudocode}
\usepackage{array}
\usepackage{booktabs}
\usepackage{multirow}
\usepackage{makecell}
\usepackage{mathtools}
\usepackage{amsthm}

\newtheorem{theorem}{Theorem}

\newcommand{\E}{\mathbb{E}}

\newcommand{\Y}{\mathcal{Y}}
\newcommand{\X}{\mathcal{X}}

\newcommand{\G}{\mathcal{G}}
\newcommand{\U}{\mathcal{U}}

\DeclareMathOperator*{\argmax}{arg\,max}

\DeclareMathOperator{\st}{start}

\begin{document}
 
\title{Low Complexity Algorithms for Transmission of Short Blocks over the BSC with Full Feedback}


\author{\IEEEauthorblockN{Amaael~Antonini, Hengjie~Yang and Richard~D.~Wesel}
\IEEEauthorblockA{
Department of Electrical and Computer Engineering\\
University of California, Los Angeles, Los Angeles, CA 90095, USA\\
Email: \{amaael, hengjie.yang, wesel\}@ucla.edu}
}


\maketitle

\begin{abstract}
Building on the work of Horstein, Shayevitz and Feder, and Naghshvar \emph{et al.}, this paper presents algorithms for low-complexity sequential transmission of a $k$-bit message over the binary symmetric channel (BSC) with full, noiseless feedback. To lower complexity, this paper shows that the initial $k$ binary  transmissions can be sent before any feedback is required and groups messages with equal posteriors to reduce the number of posterior updates from exponential in $k$ to linear in $k$. Simulation results demonstrate that achievable rates for this full, noiseless feedback system approach capacity rapidly as a function of average blocklength, faster than known finite-blocklength lower bounds on  achievable rate with noiseless active feedback and significantly faster than finite-blocklength lower bounds for a stop feedback system.
\end{abstract}

{\let\thefootnote\relax\footnote{{This research is supported by National Science Foundation (NSF) grant CCF-1955660. Any opinions, findings, and conclusions or recommendations expressed in this material are those of the author(s) and do not necessarily reflect views of the NSF.}}}

\section{Introduction}

Shannon \cite{Shannon1956} showed that feedback cannot improve the capacity of a discrete memoryless channel (DMC). However, Burnashev \cite{Burnashev1976} showed that feedback combined with variable length coding can significantly increase the exponent with which the frame error rate (FER) decreases with blocklength. Polyanskiy \emph{et al.} \cite{Polyanskiy2010,Polyanskiy2011} derived lower bounds on finite blocklength achievable rates with and without feedback that demonstrate the benefit to achievable rate of ``stop feedback,'' which is feedback that can only inform the transmitter when transmission should be terminated.


Even better performance should be attainable when stop feedback is replaced by feedback of all received symbols.   For the binary symmetric channel (BSC) with noiseless feedback, Horstein \cite{Horstein1963} presented a simple and elegant one-phase transmission scheme that uses full feedback to achieve the capacity of the BSC \cite{Shayevitz2011}.  Since Horstein's work, several authors proposed various transmission schemes for BSC with full noiseless feedback under variant settings, in order to achieve the capacity or Burnashev's optimal error exponent, e.g., \cite{Schalkwijk1971,Schalkwijk1973,Tchamkerten2002,Tchamkerten2006,Naghshvar2012}. Naghshvar \emph{et al.} \cite{Naghshvar2012,Naghshvar2015} presented a finite-blocklength version of Horstein's scheme, which they show attains the capacity and Burnashev's optimal error exponent. 

This paper focuses on the finite blocklength version of Horstein's scheme described in \cite{Naghshvar2012,Naghshvar2015}.
Horstein's scheme works as follows: For a set of $M$ messages and a given target error probability $\epsilon$, consider the unit interval initially partitioned into $M$ equal sub-intervals. Each sub-interval represents a message, and the length of each sub-interval denotes the posterior of the message. After each transmission, the receiver and transmitter (utilizing the full feedback) both use the channel output to compute new posteriors for each messages and update the sub-interval lengths accordingly.

The transmitter sends bit $0$ if the sub-interval corresponding to the true message lies entirely above the midpoint, and sends bit $1$ if it lies entirely below the midpoint. However, if the midpoint lies within the sub-interval of the true message, the transmitter sends $0$ or $1$ randomly according to the fraction of the portion of sub-interval that is above or below the midpoint. The transmission terminates when the length of the sub-interval of any message exceeds $1-\epsilon$.  Although the encoder behavior is essentially the same, we consider the {\em communication phase} to be when no message has a posterior greater than 0.5 and the {\em confirmation phase} to be when any message has a posterior greater than 0.5.  

Unlike the Horstein scheme that sets the midpoint as a hard decision threshold for the transmitter, Naghshvar \emph{et al.} \cite{Naghshvar2012,Naghshvar2015} assigns each message to one of two sets $S_0, S_1$.  The two sets must satisfy the requirement that the difference of the sums of the posteriors $P(S_0)-P(S_1)$ is less than any individual posterior in $S_0$, where we require that  $P(S_0)>P(S_1)$.  We refer to this transmission scheme as the \emph{small-enough-difference} (SED) encoder because at each transmission, the algorithm seeks a two-way partitioning with a bounded small difference. The transmitter sends a $0$ if the true message is in $S_0$, and a $1$ otherwise.

Actual implementation of the SED encoder requires significant complexity.  Perhaps for this reason, Naghshvar \emph{et al.} did not present any simulation results but rather provide bounds on how a theoretical implementation would perform.

As its primary contribution, this paper provides algorithms for transmission of short blocks (on the order of $k=300$ bits) on the BSC that can be implemented and presents simulation results.  These algorithms are made possible by three primary insights:  1) The first $k$ transmissions can be sent with partitions that achieve $P(S_0) = P(S_1)$ exactly without requiring any feedback.  2) After the initial $k$ transmissions, even though there are $2^k$ different messages, there are only $k+1$ different posterior probabilities.  Grouping messages according to their posterior probabilities significantly reduces complexity since one computation computes the posterior for all messages in the group. 3) While an SED encoder can be implemented with relatively low complexity using this grouping as a starting point, an even simpler algorithm that relaxes the requirements on the maximum difference in the probabilities of $S_0$ and $S_1$ achieves essentially the same performance.  These new implementations allow simulations that demonstrate how, for a fixed target FER, a quantization effect leads to a non-monotonic rate increase as $k$ grows.  This non-monotonic behavior can be avoided by using randomization to overcome the quantization effect.  As a final contribution, this paper shows how achievable rate changes as a function of target FER.

The rest of this paper is organized as follows: Sec. \ref{sec:NewAlgoithmBasics} presents the system model and two tools used in the initial operation of the new algorithms.  Sec. \ref{sec:First_k_no_feedback} shows that the first $k$ transmissions can be sent before any feedback is required.  Sec. \ref{sec:Hamming_distance_to_received_message} presents a technique of ordering and labeling possible messages according to their Hamming distance from the initial $k$ received bits.  Sec. \ref{sec:Sorting} describes how the messages that are ordered and labeled as in Sec. \ref{sec:NewAlgoithmBasics} can be sorted and partitioned into two sets that either meet the SED criterion of \cite{Naghshvar2012, Naghshvar2015} or a more relaxed criterion that requires only one split of a labeled group of equal-posterior messages per transmission.  Simulation results show that the relaxed criterion has a negligible effect on the rate as compared to the SED criterion.  Sec. \ref{sec:RepairingJaggedFERs} uses the threshold randomization to mitigate the rate penalty incurred for some small values of $k$ when integer thresholds significantly exceed the required FER performance. Sec. \ref{sec:ratevsFER} explores the tradeoff between FER and rate, and  Sec. \ref{sec:conclusions} concludes the paper.

\section{Initial Transmission and Labeling}
\label{sec:NewAlgoithmBasics}

\begin{figure}[t]
\centering
\includegraphics[width=0.45\textwidth]{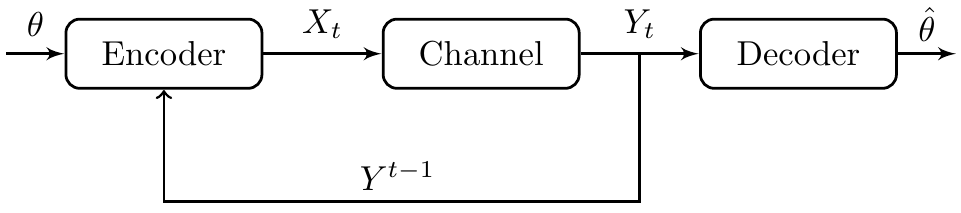}
\caption{System diagram of a DMC with full, noiseless feedback.}
\label{fig: system model}
\end{figure}

\subsection{System Model and the SED Encoder}
\label{sec: system model}
The basic system model is depicted in Fig. \ref{fig: system model}, in which the forward channel is a DMC described by an ordered triple $(\X,\Y, P(Y|X))$ and the feedback channel noiselessly provides the received channel outputs to the receiver. Let $\theta$ be the true message uniformly drawn from a message set $\Omega=\{1,2,\dots,M\}$. At each time instant $t$, $t=1,2,\dots$, the transmitter is aware of both the true message $\theta$ and the received symbols $Y^{t-1}=(Y_1,Y_2,\dots,Y_{t-1})$, thanks to the noiseless feedback. The total transmission time (or the number of channel uses, or blocklength) $\tau$ is a random variable that is governed by a stopping rule that is a function of the observed channel outputs.

In order to communicate $\theta$ from the transmitter to the receiver, the transmitter produces channel inputs $X_t$, $t=1,2,\dots,\tau$, as a function of $\theta$ and $Y^{t-1}$, i.e.,
\begin{align}
    X_t=e_t(\theta,Y^{t-1}), \quad t=1,2,\dots,\tau,
\end{align}
for some encoding function $e_t:\Omega\times\Y^{t-1}\to\X$. After observing $\tau$ channel outputs $Y^\tau$, the receiver makes a final estimate $\hat{\theta}$ of the true message $\theta$, which is a function of $Y^\tau$, i.e.,
\begin{align}
    \hat{\theta}=d(Y^\tau),
\end{align}
for some decoding function $d: \Y^\tau\to \Omega$.
An error occurs if $\hat{\theta}\ne \theta$ and the probability of error is given by $P_e=\Pr\{\theta\ne \hat{\theta}\}$.

For a given target error probability $\epsilon$, $\epsilon>0$, the fundamental problem of variable-length coding is to design the encoding function $e_t(\cdot)$, decoding function $d(\cdot)$, a stopping rule that defines the stopping time $\tau$, such that $P_e\le\epsilon$ and the average blocklength $\E[\tau]$ is minimized.

In \cite{Naghshvar2012} and \cite{Naghshvar2015}, Naghshvar \emph{et al.} considered the following encoding rule (called the SED encoder), the decoding rule, and the stopping rule for the BSC$(p)$, $0<p<1/2$.

\textbf{The SED encoding rule}: at each time $t$, $t=1,2,\dots,\tau$, with the full, noiseless feedback $Y^{t-1}$, the transmitter considers the \emph{belief state} $\bm{\rho}(t)$ at time $t$
\begin{align}
    \bm{\rho}(t)=[\rho_1(t), \rho_2(t), \dots, \rho_M(t)],
\end{align}
where
\begin{align}
    \rho_i(t)\triangleq \Pr\{\theta=i|Y^{t-1}\},
\end{align}
with the convention that $\rho_i(1)=1/M$.
Using Bayes rule, $\bm{\rho}(t+1)$ can be updated recursively from $\bm{\rho}(t)$ upon receiving $y_t$, i.e.,
\begin{align}
    \rho_i(t+1)=\frac{\rho_i(t)P(Y=y_{t}|X=e_{t}(i,Y^{t-1}))}{\sum_{j\in\Omega}\rho_j(t)P(Y=y_{t}|X=e_{t}(j,Y^{t-1}))} \label{eq: Bayes's rule}
\end{align}

Next, the transmitter partitions $\Omega$ into two subsets $S_0(t)$ and $S_1(t)$ such that
\begin{align}
0\le\sum_{\mathclap{i\in S_0(t)}}\ \rho_i(t)-\sum_{\mathclap{i\in S_1(t)}}\ \rho_i(t)\le\min_{\mathclap{i\in S_0(t)}}\ \rho(t). \label{eq: SED encoder}
\end{align}
Then, $X_t=0$ if $\theta\in S_0(t)$ and $X_t=1$ otherwise.

\textbf{The stopping rule and decoding rule}: the stopping time $\tau$ and the estimate $\hat{\theta}$ are given by
\begin{align}
    \tau=&\min\{t: \max_{i\in\Omega}\rho_i(t)\ge1-\epsilon\}\label{eq: stopping rule}\\
\hat{\theta}=&\argmax_{i\in\Omega}\rho_i(\tau).\label{eq: estimate message}
\end{align}
Clearly, the probability of error under stopping rule \eqref{eq: stopping rule} meets the desired constraint,
\begin{align}
P_e=\E[1-\max_{i\in\Omega}\rho_i(\tau)]\le\epsilon.
\end{align}

We remark that if $M=2^k$, $k=1,2,\dots$, the partitioning algorithm for the SED encoder described in Naghshvar \emph{et al.} \cite{Naghshvar2012,Naghshvar2015} requires  exponential complexity in $k$, making it difficult to implement in practice. Thus, a low complexity partitioning algorithm that can still guarantee a similar or equal performance as the SED encoder is desired.

\subsection {Sending the First $k$ Transmissions without Feedback}
\label{sec:First_k_no_feedback}
Consider the BSC$(p)$, $0<p<1/2$, define $q=1-p$, denote message $i \in \Omega = \{1,...,M\}, M= 2^k$ by its binary representation $\bm{b}^{(i)}=(b_0^{(i)},b_1^{(i)},\dots,b_{k-1}^{(i)})_2$ and define the posteriors of $S_0(t)$ and $S_1(t)$ given $Y^{t-1}$ after partitioning $\Omega$ at time $t$ by:
\begin{align}
    \pi_x(t)=\sum_{{i\in S_x(t)}}\ \rho_i(t),\quad x\in\{0,1\}.
\end{align}
Also, let the posterior updates after transmission $t$ be:

$P(S_0(t)|Y_{t},Y^{t-1})= w_{0,t}\pi_0(t)$, $P(S_1(t)|Y_{t},Y^{t-1})=w_{1,t} \pi_1(t)$ where the weights $w_{0,t}$, $w_{1,t}$ are given by:
\begin{align}
w_{0,t} \triangleq
\begin{cases}
      \frac{q}{q\pi_0(t) + p\pi_1(t)}, & \text{if } y_{t} = 0 \\
      \frac{p}{p\pi_0(t) + q\pi_1(t)}, & \text{if } y_{t} = 1                
\end{cases}   \\   
w_{1,t} \triangleq
\begin{cases}
      \frac{p}{q\pi_0(t) + p\pi_1(t)}, & \text{if } y_{t} = 0 \\
      \frac{q}{p\pi_0(t) + q\pi_1(t)}, & \text{if } y_{t} = 1
\end{cases} 
\end{align}

Note that $\forall i \in S_x(t), \ \rho_i(t+1) = \rho_i(t) \cdot w_{x,t} $,  
and if $\pi_0(t) = \pi_1(t) = \frac{1}{2}$, then:
\begin{align}
(w_{0,t},w_{1,t}) = 
\begin{cases}
      (2q, 2p), & \text{if } y_{t} = 0 \\
      (2p, 2q), & \text{if } y_{t} = 1  
\end{cases}      
\end{align}
\begin{theorem}\label{theorem: equal partitions before k transmisisons}
Let $\theta$ be a $k$-bit message uniformly drawn from $\Omega=\{1,2,\dots,M\}$, $M=2^k$. Then, for $t\le k$, there is a systematic method to partition $\Omega$ into $S_0(t)$ and $S_1(t)$ such that $\pi_0(t) = \pi_1(t) = \frac{1}{2}$, that is independent of the transmitted and received sequences $X^t$, $Y^t$. 
\end{theorem}


\begin{IEEEproof}
We show that the systematic partitioning rule
    \begin{align}
    S_0(t) = \{i \in \Omega: ~ b^{i}_{t-1} = 0\}  \label{eq:S0}\\
    S_1(t) = \{i \in \Omega: ~ b^{i}_{t-1} = 1\} \label{eq:S1}
\end{align}
yields  $\pi_0(t) = \pi_1(t) = \frac{1}{2}$ for $t\le k$.
For all times $t<k$, the partitioning rule (\ref{eq:S0}-\ref{eq:S1}) does not consider the final $k-t$ bits of the message, i.e. $b_t^{(i)},b_{t+1}^{(i)},\dots,b_{k-1}^{(i)}$. Therefore, the $2^{k-t}$ messages that share the first $t$ bits sequence are assigned together to the same set in each of the first $t$ partitionings.  Thus each of these $2^{k-t}$ messages has the same posterior at time $t$. At time  $t+1\le k$ each group of $2^{k-t}$ equal-posterior messages is  split by (\ref{eq:S0}-\ref{eq:S1}) into two groups of $2^{k-t-1}$ messages with equal posteriors, one group with $b_t = 0$ and the other with $b_t = 1$.  These groups are assigned to $S_0(t+1)$ and $S_1(t+1)$ respectively, resulting in $\pi_0(t+1) = \pi_1(t+1) = \frac{1}{2}$.
\end{IEEEproof}

An immediate consequence of Theorem \ref{theorem: equal partitions before k transmisisons} is that we can transmit the first $k$ bits systematically while maintaining SED condition of \eqref{eq: SED encoder} in the first $k$ transmissions even without feedback. That is, if the binary representation of $\theta$ is $\bm{b}^{(\theta)}=(b_0^{(\theta)},b_1^{(\theta)},\dots,b_{k-1}^{(\theta)})_2$, $b_i^{(\theta)}\in\{0,1\}$, $i=0,1,\dots,k-1$, then $X_t=b^{(\theta)}_{t-1}$ is always possible as long as $t\le k$, since we can always label the subset including $\theta$ by $b^{(\theta)}_{t-1}$.

\subsection {Ordering and Labeling Possible Messages}
\label{sec:Hamming_distance_to_received_message}
After transmitting the first $k$ bits systematically, the receiver possesses a noisy version $y^k=(y_1,y_2,\dots,y_k)$ of the $k$-bit true message $\theta$ over the BSC$(p)$ and the transmitter is aware of the received bits thanks to the noiseless, full feedback. 

First, we note that, after the $k$-th transmission, the posterior of each message $i\in\Omega$ can be explicitly computed according to the Hamming distance to the received sequence $y^k$. 
Thus, if the Hamming distance between $\bm{b}^{(j)}$ and $y^k$ is $d_H(\bm{b}^{(j)},y^k)=d_{i,y^k}$, the posterior of message $j\in\Omega$, after the $k$-th transmission is given by
\begin{align}
    \rho_j(k+1)=p^{d_{i,y^k}}q^{(k-d_{i,y^k})}.
\end{align}
Thus, each message with distance $d_{j,y^k}$ can be categorized into one of $(k+1)$ \emph{groups} $\G_d(k)$, $d=0,1,\dots,k$, with group $\G_d(k)$ having the same posterior $p^dq^{k-d}$. The cardinality of group $\G_d(k)$ after the $k$-th transmission is given by $\binom{k}{d}$. If we introduce the lexicographical ordering for each group, then there is a one-to-one correspondence between message $\bm{b}$ in $\G_{d}(k)$ to an index, that we denote by $ n_{d}(\bm{b})$ and define in next paragraph, which later greatly simplifies the group split and list merge operations.

Next, we show that the index $n_{d}(\bm{b})$ can be calculated efficiently, which has been proposed and studied in the context of enumerative source coding \cite{Cover1973}. For completeness of this paper, we introduce it in what follows. In general, consider the function
\begin{align}
    n_d(\bm{b}): \U_d \to \{0,1,\dots,\binom{k}{d}-1\},
\end{align}
where $\U_d=\{\bm{b}\in\{0,1\}^k, w_H(\bm{b},y^k)=d\}$ consists of all messages whose binary representation is of distance $d$. Let $0\le i_1<i_2<\dots<i_{d}\le k-1$ denote the position of $1$'s for message $\bm{b}$. Thus, $n_d(\bm{b})$ is given by
\begin{align}
    n_d(\bm{b})=\binom{i_1}{1}+\binom{i_2}{2}+\cdots+\binom{i_d}{d}.
\end{align}
Conversely, given $n_d(\bm{b})$, we can easily recover message $j$ by sequentially determining $i_d, i_{d-1},\dots,i_1$. Namely, $i_d$ is determined by the largest integer such that $\binom{i_d}{d}\le n_d(j)$; next, $i_{d-1}$ is determined by the largest integer such that $\binom{i_{d-1}}{d-1}\le n_d(j)-\binom{i_d}{d}$; so on and so forth. 

Hence, each group $\G_j(t)$ can be compactly described by an ordered tuple
\begin{align}
    \G_j(t)=\big(d,n_{\st}, N, \delta\big)
\end{align}
where $d$ is the Hamming distance from $y^k$, $n_{\st}$ is the index of the first element, $N$ is the total elements in $\G_j(t)$ and $\delta$ is the posterior associated with $\G_j(t)$. For example, after the $k$-th transmission, $j=0,1,\dots,k$,
\begin{align}
    \G_j(k+1)=\big(d,n_{\st}, N, \delta \big)=\big(j, 0,\binom{k}{j},p^dq^{k-j}\big).
\end{align}
The number of groups at time $t$, $t\ge k$, depends on the partitioning algorithm and $Y^t$, if no group is split, then the number of groups remains $k+1$ over time.

\section{Sorting, Grouping, and Splitting Posteriors}
\label{sec:Sorting}

We propose a system that transmits $\bm{b}^{(\theta)}$ in the first $k$ transmissions. After the $k$-th transmission, the transmitter first generates a list of $(k+1)$ groups $\G_j(k+1)=(d,n_{\st},N,\delta)$, $d=0,1,\dots,k$ in the order of decreasing posteriors $\delta$.

At the $t$-th transmission, $t>k$, the transmitter aims at partitioning $\Omega$ into two subsets $S_0(t)$, $S_1(t)$, by only using \emph{group movement} and \emph{group split} operations. Assume that the group $\G(t)=\{d,n_{\st},N,\delta\}$ is to be split at $(n_{\st}+N_1)$-th position, $N_1\in\{1,2,\dots,N-1\}$. The resultant two subgroups are readily given by
\begin{align}
    \G^{(1)}(t)=&\big(d,n_{\st},N_1,\delta\big),\\
    \G^{(2)}(t)=&\big(d,n_{\st}+N_1,N-N_1,\delta\big).
\end{align}
After the $t$-th transmission, we update the posteriors by updating the associated posterior in each group. For example, if $S_0(t)$ is boosted by $w_{0,t}$ and $S_1(t)$ is attenuated by $w_{1,t}$, then the groups in $S_0(t), S_1(t)$ are updated to
\begin{align}
    \G(t)=\big(d,n_{\st},N,w_{0,t}\delta \big),\quad \text{if }\G\in S_0(t)\\
    \G(t)=\big(d,n_{\st},N,w_{1,t}\delta \big),\quad \text{if }\G\in S_1(t).
\end{align}

\subsection{Achieving the Small-Enough-Difference Criterion}
\label{sec:SED} 
In order to achieve optimal partitioning of the list into $S_0(t)$ and $S_1(t)$, the two new lists need to meet the SED criterion of \cite{Naghshvar2012, Naghshvar2015} given by \eqref{eq: SED encoder}. We implement Algorithm II \cite{Naghshvar2012, Naghshvar2015} in an equivalent way, with one modification. The equivalent method is to assign the whole list to $S_1(t)$ first instead of $S_0(t)$, and move the message with largest probability to $S_0(t)$ instead of the message with smallest probability to $S_1(t)$. When we first have that next message assigned to $S_0(t)$ will cause $\pi_0(t) \ge 0.5$, which might require splitting one group, if we cannot meet SED criterion, instead of swapping the list, we test if the whole or a splitting of the next group would be enough to meet SED criterion, and use it if it does, else proceed as we would have otherwise.

\subsection{Reconstructing the Decoded Message}
\label{Reconstruction}
Once the confirmation phase is finished, a unique group $\hat{\G(t)}=(d,n_{\st},N,\delta)$ contains a single error case for which we need to determine the decoded message $\hat{\theta}$, i.e., $N=1$, $\delta\ge1-\epsilon$. This is accomplished by the inverse of $n_d(\bm{b})$ as discussed in Sec. \ref{sec:Hamming_distance_to_received_message}.


\subsection{A Relaxed Criterion that Minimizes Splits}
\label{Relaxed}
\begin{figure}[t]
\centering
\includegraphics[width=0.47\textwidth]{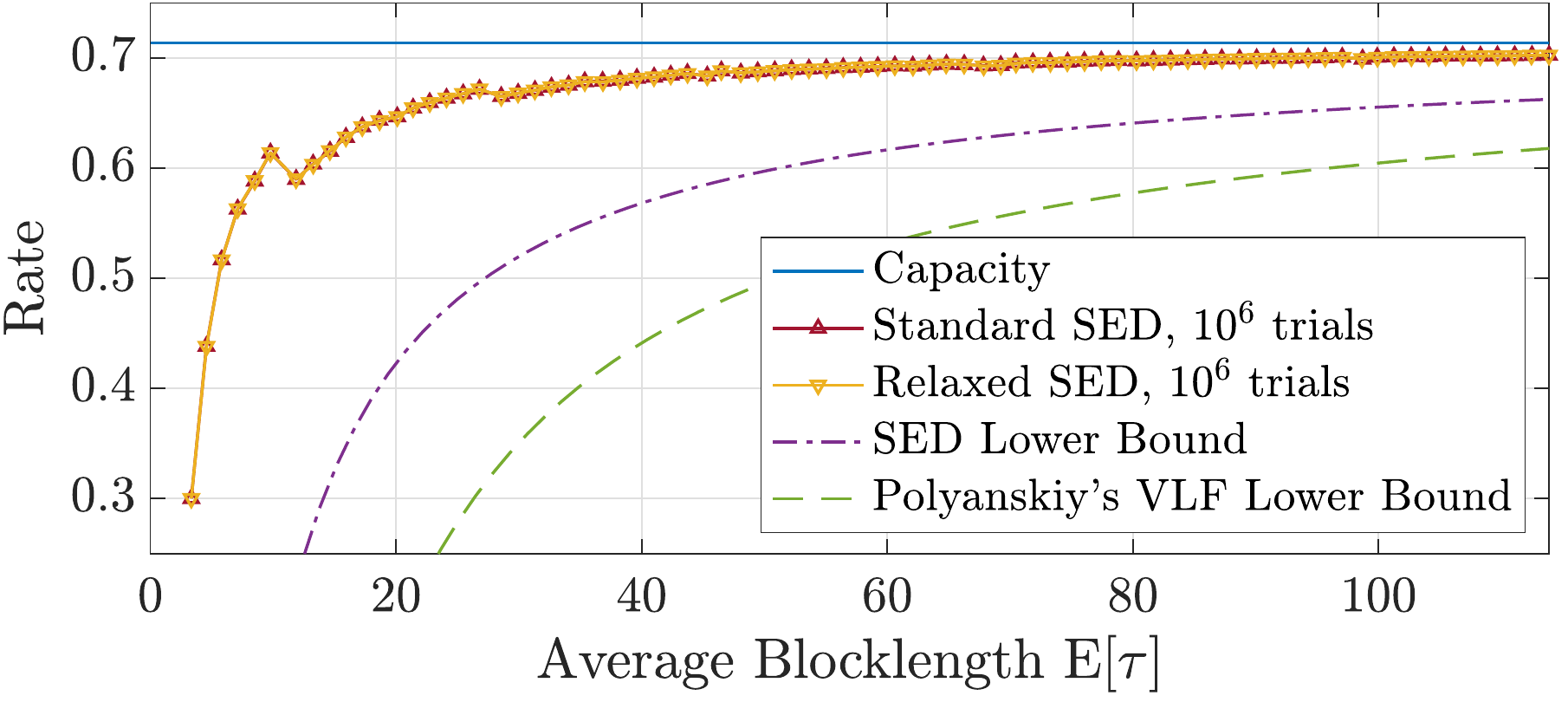}
\caption{Rate as a function of average blocklength for two algorithms, over the BSC$(0.05)$ with full, noiseless feedback. One algorithm achieves the SED criterion, and the other algorithm achieves a relaxed criterion that requires at most one split of a group of messages per transmission.  Also shown are Polyanskiy's VLF lower bound for stop feedback and the SED lower bound from \cite{Yang2019} $\epsilon=10^{-3}$. }
\label{fig: rate_vs_blocklength}
\end{figure}

We next evaluate the performance of a system that limits the number of group splits to a maximum of one per transmission, which might prevent the system from meeting the SED criterion. But it guarantees a relaxed version of SED criterion given by:
\begin{align}
0\le\sum_{\mathclap{i\in S_0(t-1)}}\ \rho_i(t-1)-\sum_{\mathclap{i\in S_1(t-1)}}\ \rho_i(t-1)\le2\min_{\mathclap{i\in S_0(t-1)}}\ \rho(t-1).
\end{align}
To implement this relaxed condition, we start the procedure as before, moving messages into $S_0(t)$ in descending order. We continue until one message with posterior $\rho(t)$ is moved into $S_0(t)$ such that $\pi_0(t)\ge0.5$. Since the last movement yields  $\pi_0(t)\le 0.5+\rho(t)$, we conclude $\pi_1(t) > 0.5-\rho(t)$ and thus, $\pi_0(t)-\pi_1(t)\le 2\rho(t)$. Note that $\rho(t)$ is the smallest posterior in $S_0$, hence, the relaxed criterion is met.
Fig. \ref{fig: rate_vs_blocklength} shows that the relaxed criterion which requires at most a single split per transmitted bit exhibits an indistinguishable performance, compared to the original SED encoder.

\subsection{System Complexity}
\label{Complexity}
The system's complexity can be characterized by the number of transmissions and the operations at each transmission. The number of transmissions is bounded by a linear function of $k$, as the scheme approaches capacity asymptotically. The operations at each transmission $t$ during the communication phase are posterior updates, merging of $S_0(t-1)$, $S_1(t-1)$ into an ordered list and partition of the new list into $S_0(t)$ and $S_1(t)$. These operations are linear functions of the list size. The list grows as group splitting operations are performed. When a split is required, the probability that it is final can be estimated at $0.5$, and therefore the average number of splits at each partition is close to $2$. Then, the list size is a linear function of $k$ and the system complexity is of order $\mathcal{O}(k^2)$.
The memory requirement for this system depends on the length of the list, linear in $k$. We greatly simplify calculation by storing a triangular array of combinations size $\frac{(k+1)k}{2}$. The storage requirement is then also of order $\mathcal{O}(k^2)$.

\section{Randomization, Grouping, FER vs. Rate}
\label{sec:RepairingJaggedFERs}
This section explores how randomization can remove the ``notch'' in rate visible in Fig. \ref{fig: rate_vs_blocklength} at $k=11$, how condensing a large group of small-probability messages can further reduce complexity, and how the choice of target FER affects achievable rate.

\subsection{Stopping Threshold Randomization}
 The degree to which the posterior exceeds $1-\epsilon$ at the conclusion of the confirmation phase depends on a threshold that is effectively an integer describing the required difference between number of bits received in the confirmation phase that confirm the candidate message and the number that contradict it. The notch occurs because the integer threshold causes the posterior to far exceed $1-\epsilon$, achieving an FER well below $\epsilon$.  This extra reliability incurs a rate penalty that induces the notch in  Fig. \ref{fig: rate_vs_blocklength}.

\begin{figure}[t]
\centering
\includegraphics[width=0.47\textwidth]{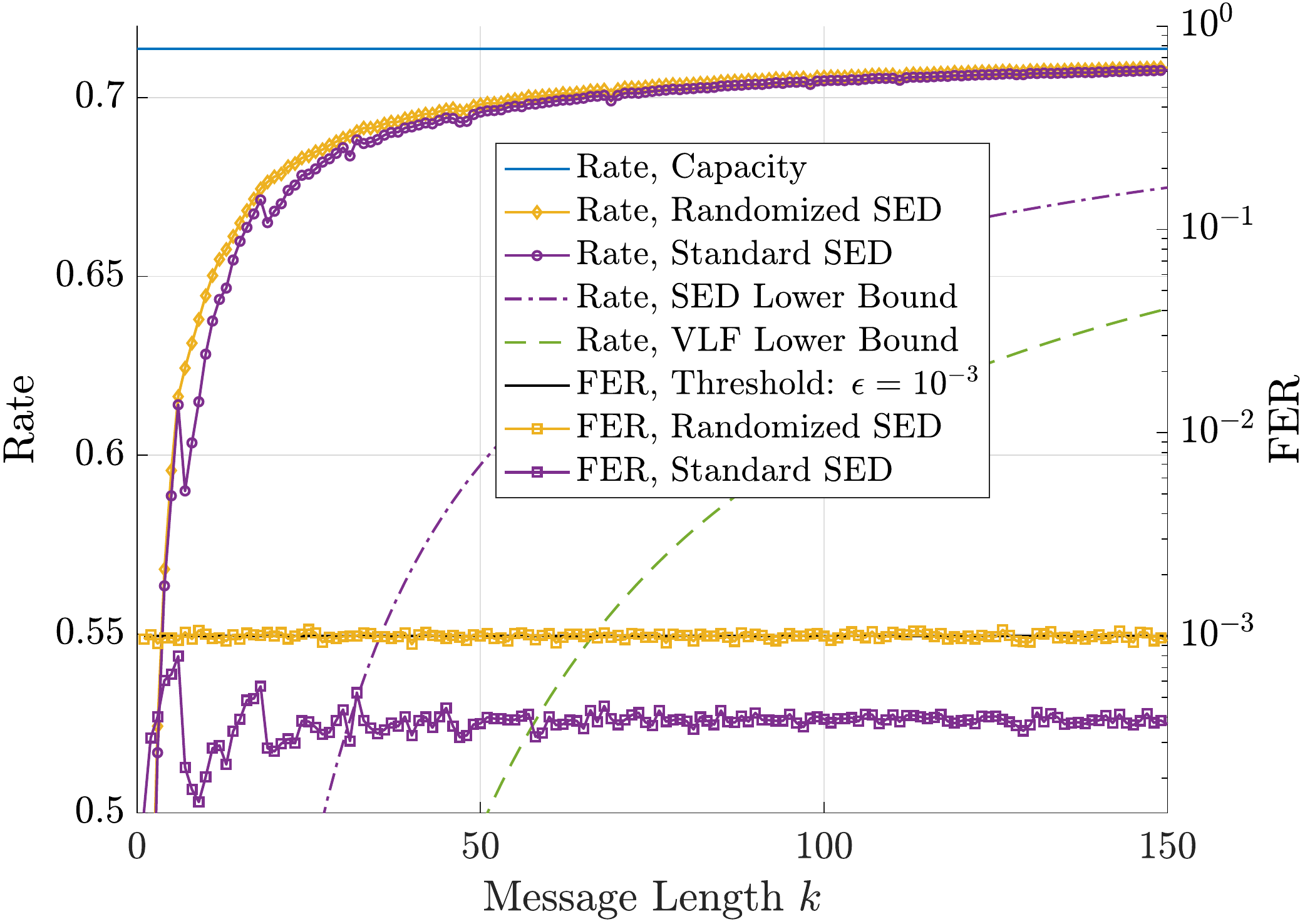}
\caption{Left axis: Rate vs. average blocklength over the BSC$(0.05)$ for $10^6$ trials of SED with a fixed threshold that guarantees $\text{FER} <10^{-3}$ and a randomized threshold that closely approximates FER of $10^{-3}$. Right axis: the corresponding FERs achieved by the the two stopping criteria.}
\label{fig:rate_with_randomization}
\end{figure}

Threshold randomization removes this rate loss. Fig. \ref{fig:rate_with_randomization} shows the rate achieved by the standard SED algorithm and the smoother and higher rate curve achieved by randomly selecting a threshold between the standard integer value and a threshold that is the next smaller integer. Each of the two thresholds has a corresponding posterior at termination, and the randomization is biased  ensure that the expetced posterior is the target FER.  Fig. \ref{fig:rate_with_randomization} also shows the corresponding FERs of the standard and randomized approaches. 

\subsection{Grouping Messages to Reduce Complexity}
\begin{figure}[t]
\centering
\includegraphics[width=0.47\textwidth]{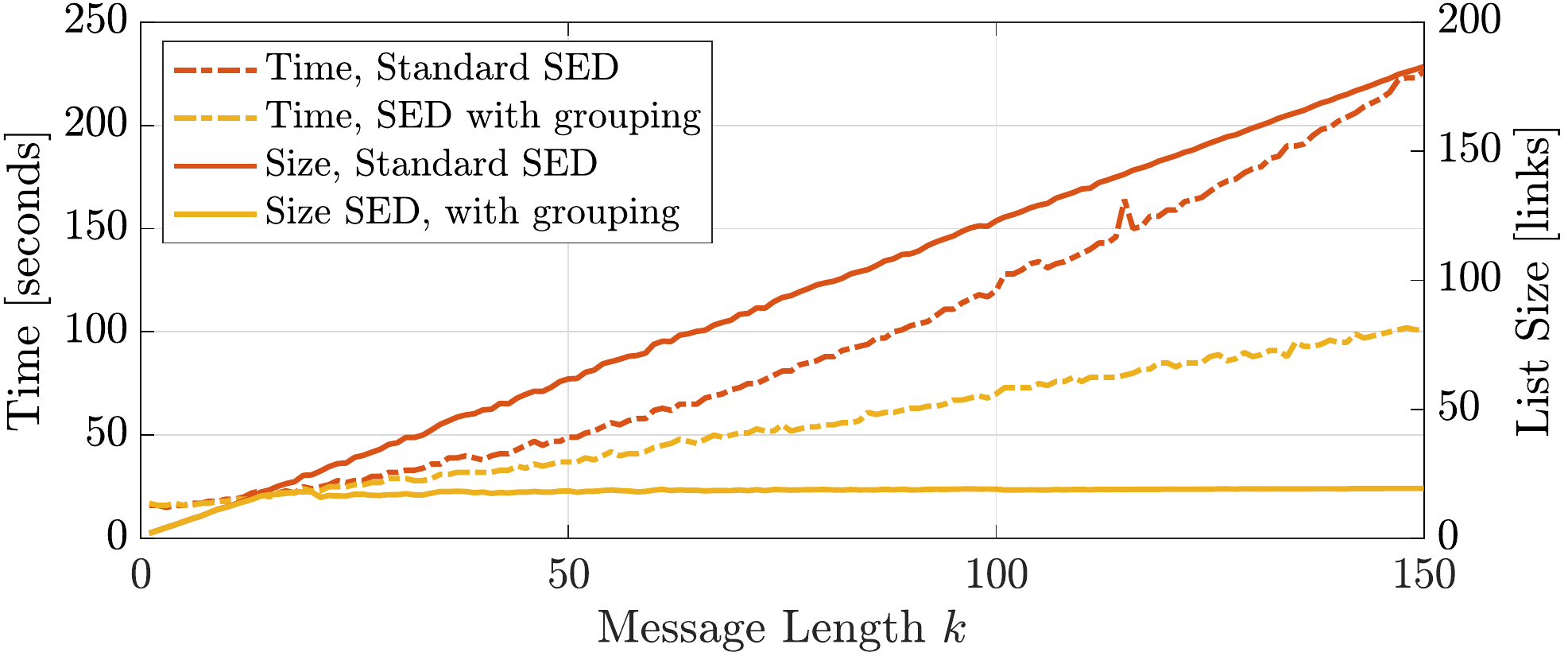}
\caption{Right: Average list size and left: time to process $10^5$ transmissions as a function of message size $k$ over BSC$(0.05)$ for the SED criterion and the relaxation criterion of \ref{grouping}. 
}
\label{fig:complexity}
\end{figure}

From the relaxation in \ref{Relaxed}, the front of the list is assigned to $S_0(t)$ and the rest to $S_1(t)$. Hence, a large number of groups at the back of the list are consistently assigned to $S_1(t)$ because their accumulated probability rarely grows large enough. For small crossover probability $p < 0.25$ the majority of the groups are consistently assigned to $S_1(t)$. We separate these groups into a list denoted by $S_{\text{tail}}(t)$, and only track its accumulated probability $P(S_{\text{tail}}(t))$ and the common weight update that we denoted $w_{\text{tail}}$. We merge its links into the main list when $P(S_{\text{tail}}(t))$ crosses a threshold, therefore, no messages are lost and there is no performance degradation. This way, for $k$ in the range of $400$ the size of the list is dominated by thresholds to remove and recover links into and from $S_{\text{tail}}(t)$ rather than $k$. The average list size becomes constant rather than linear in $k$ reducing the total time complexity from $\mathcal{O}(k^2)$ to $\mathcal{O}(k)$, as shown in Fig. \ref{fig:complexity}.  This allows efficient transmission of messages with larger values of $k$.
\label{grouping}

\begin{figure}[t]
\centering
\includegraphics[width=0.47\textwidth]{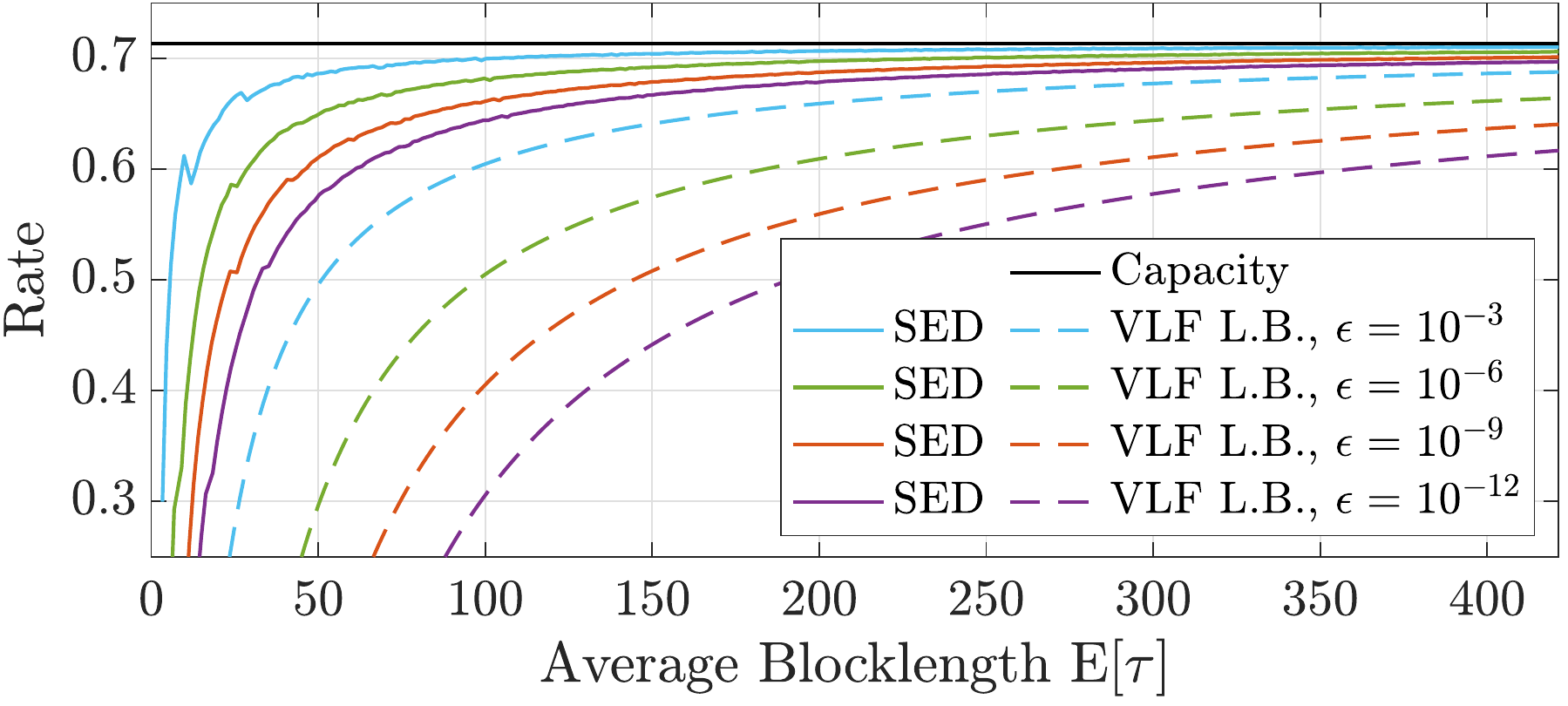}
\caption{Rate as a function of blocklength for the SED relaxed criterion algorithm, that limits list size, implemented with $10^5$ trials and four thresholds that guarantee respectively FERs of  $10^{-3}$, $10^{-6}$, $10^{-9}$, and $10^{-12}$ over the BSC$(0.05)$ with full, noiseless feedback. Also shown the corresponding Polyanskiy's VLF lower bounds for stop feedback.}
\label{fig:ratesfor4FERs}
\end{figure}

\subsection{ The Tradeoff Between FER and Rate}
\label{sec:ratevsFER}
Fig. \ref{fig:ratesfor4FERs} shows how increasing the reliability requirement affects the rate performance as a function of the target FER for the standard SED algorithm using a non-randomized threshold.   Fig. \ref{fig:ratesfor4FERs} uses only $10^5$ trials to produce the rate curves even for FERs as low as $10^{-12}$.  However, note that the FER target is necessarily achieved by the SED threshold 
and $10^5$ trials is more than sufficient to estimate {\em rate} \cite{Williamson2015}.  For an average blocklength of 400 transmitted symbols, the rate achieved by SED is similar for the entire range of FERs considered; there is little rate penalty in requiring an FER of $10^{-12}$.  In contrast, the VLF lower bound on achievable rate with stop feedback from \cite{Polyanskiy2011} shows a noticeable penalty to achieve FER of $10^{-12}$ with 400 transmissions. The VLF lower bound is similar to the simulated SED performance for FER of $10^{-3}$, which itself is surprising given that the stop feedback of VLF is far more constrained than the full feedback used by SED.

\section{Conclusions}

This paper introduces algorithms for low-complexity sequential transmission of a $k$-bit message over the binary symmetric channel (BSC) with full, noiseless feedback. The initial $k$ binary  transmissions can be sent before any feedback is required.   A technique for managing posterior updates by grouping messages with equal-value posteriors lowers complexity.  Relaxing the SED criterion further lowers complexity without sacrificing performance.  Threshold randomization avoids the rate penalty incurred by integer thresholds that force an FER well below the target.  Simulation results agree with the SED lower bound of \cite{Yang2019} and show the trade-off of rate vs. target FER.

\label{sec:conclusions}

\bibliographystyle{IEEEtran}
\bibliography{IEEEabrv,references}

\begin{thebibliography}{10}
\providecommand{\url}[1]{#1}
\csname url@samestyle\endcsname
\providecommand{\newblock}{\relax}
\providecommand{\bibinfo}[2]{#2}
\providecommand{\BIBentrySTDinterwordspacing}{\spaceskip=0pt\relax}
\providecommand{\BIBentryALTinterwordstretchfactor}{4}
\providecommand{\BIBentryALTinterwordspacing}{\spaceskip=\fontdimen2\font plus
\BIBentryALTinterwordstretchfactor\fontdimen3\font minus
  \fontdimen4\font\relax}
\providecommand{\BIBforeignlanguage}[2]{{%
\expandafter\ifx\csname l@#1\endcsname\relax
\typeout{** WARNING: IEEEtran.bst: No hyphenation pattern has been}%
\typeout{** loaded for the language `#1'. Using the pattern for}%
\typeout{** the default language instead.}%
\else
\language=\csname l@#1\endcsname
\fi
#2}}
\providecommand{\BIBdecl}{\relax}
\BIBdecl

\bibitem{Shannon1956}
C.~Shannon, ``The zero error capacity of a noisy channel,'' \emph{IRE Trans.
  Inf. Theory}, vol.~2, no.~3, pp. 8--19, September 1956.

\bibitem{Burnashev1976}
M.~V. Burnashev, ``Data transmission over a discrete channel with feedback.
  random transmission time,'' \emph{Problemy Peredachi Inf.}, vol.~12, no.~4,
  pp. 10--30, 1976.

\bibitem{Polyanskiy2010}
Y.~Polyanskiy, H.~V. Poor, and S.~Verdu, ``Channel coding rate in the finite
  blocklength regime,'' \emph{{IEEE} Trans. Inf. Theory}, vol.~56, no.~5, pp.
  2307--2359, May 2010.

\bibitem{Polyanskiy2011}
------, ``Feedback in the non-asymptotic regime,'' \emph{{IEEE} Trans. Inf.
  Theory}, vol.~57, no.~8, pp. 4903--4925, Aug 2011.

\bibitem{Horstein1963}
M.~Horstein, ``Sequential transmission using noiseless feedback,'' \emph{{IEEE}
  Trans. Inf. Theory}, vol.~9, no.~3, pp. 136--143, July 1963.

\bibitem{Shayevitz2011}
O.~Shayevitz and M.~Feder, ``Optimal feedback communication via posterior
  matching,'' \emph{{IEEE} Trans. Inf. Theory}, vol.~57, no.~3, pp. 1186--1222,
  March 2011.

\bibitem{Schalkwijk1971}
J.~Schalkwijk, ``A class of simple and optimal strategies for block coding on
  the binary symmetric channel with noiseless feedback,'' \emph{{IEEE} Trans.
  Inf. Theory}, vol.~17, no.~3, pp. 283--287, May 1971.

\bibitem{Schalkwijk1973}
J.~Schalkwijk and K.~Post, ``On the error probability for a class of binary
  recursive feedback strategies,'' \emph{{IEEE} Trans. Inf. Theory}, vol.~19,
  no.~4, pp. 498--511, July 1973.

\bibitem{Tchamkerten2002}
A.~Tchamkerten and E.~Telatar, ``A feedback strategy for binary symmetric
  channels,'' in \emph{Proc. IEEE Int. Symp. Inf. Theory}, June 2002, pp.
  362--362.

\bibitem{Tchamkerten2006}
A.~Tchamkerten and I.~E. Telatar, ``Variable length coding over an unknown
  channel,'' \emph{{IEEE} Trans. Inf. Theory}, vol.~52, no.~5, pp. 2126--2145,
  May 2006.

\bibitem{Naghshvar2012}
M.~{Naghshvar}, M.~{Wigger}, and T.~{Javidi}, ``Optimal reliability over a
  class of binary-input channels with feedback,'' in \emph{{IEEE} Trans. Inf.
  Theory}, Sep. 2012, pp. 391--395.

\bibitem{Naghshvar2015}
M.~Naghshvar, T.~Javidi, and M.~Wigger, ``Extrinsic {J}ensen--{S}hannon
  divergence: Applications to variable-length coding,'' \emph{{IEEE} Trans.
  Inf. Theory}, vol.~61, no.~4, pp. 2148--2164, April 2015.

\bibitem{Cover1973}
T.~{Cover}, ``Enumerative source encoding,'' \emph{{IEEE} Trans. Inf. Theory},
  vol.~19, no.~1, pp. 73--77, January 1973.

\bibitem{Yang2019}
\BIBentryALTinterwordspacing
H.~Yang and R.~D. Wesel, ``Finite-blocklength performance of sequential
  transmission over bsc with noiseless feedback.'' [Online]. Available:
  \url{http://arxiv.org/abs/1902.00593}
\BIBentrySTDinterwordspacing

\bibitem{Williamson2015}
A.~R. Williamson, T.~Chen, and R.~D. Wesel, ``Variable-length convolutional
  coding for short blocklengths with decision feedback,'' \emph{{IEEE} Trans.
  Inf. Theory}, vol.~63, no.~7, pp. 2389--2403, July 2015.

\end{thebibliography}

\end{document}